\documentstyle[12pt]{article}
\begin{document}

\setlength{\textheight}{240mm}
\voffset=-25mm
\baselineskip=20pt plus 2pt
\begin{center}

{\large \bf Energy associated with a static spherically symmetric 
nonsingular black hole }\\
\vspace{5mm}
\vspace{5mm}
I-Ching Yang$^{\dagger}$ \footnote{E-mail:icyang@dirac.phys.ncku.edu.tw}
and Irina Radinschi$^{\ddagger}$ \footnote{E-mail:iradinsc@phys.tuiasi.ro}

$^{\dagger}$Department of Natural Science Education, \\
National Taitung Teachers College, \\
Taitung, Taiwan 950, Republic of China \\ 
and \\
$^{\ddagger}$Department of Physics, ``Gh. Asachi" Technical University, \\
Iasi, 6600, Romania

\end{center}
\vspace{5mm}

\begin{center}
{\bf ABSTRACT}
\end{center}
We evaluate the energy distributions of the Dymnikova space-time
using the Weinberg, Papapetrou, and M{\o}ller energy-momentum 
complexes. This result sustain the importance of the energy-momentum
complexes in the evaluation of the energy distribution of a given
space-time. To compare the energy distributions obtained by using 
several definitions, these results show that the Einstein, Tolman,
and Weinberg energy complexes are the same in Schwarzschild  
Cartesian coordinates, but the Papapetrou and the M{\o}ller are not.

\vspace{2mm}
\noindent
{PACS No.:04.20.-q, 04.50.+h}
\newpage

A large number of definitions of the gravitational energy according
to General Relativity have been given.  However, early 
energy-momentum investigations for gravitating system gave 
reference-frame-dependent energy-momentum complex~\cite{1,2,3,4}; later the 
quasilocal idea~\cite{5} was developed.  In their latest article, 
Chang, Nester and Chen~\cite{6} showed that the energy-momentum 
complexes are actually quasilocal and associated with the legitimate 
Hamiltonian boundary term. Whereas, several examples of particular 
space-times (the Kerr-Newman, the Einstein-Rosen, the Bonnor-Vaidya 
and the Kerr-Schild class) of black holes have been studied and 
different energy-momentum complexes are given the same energy 
distribution for a given space-time~\cite{7,8,9}.  Recently, 
Yang~\cite{10} employing the Einstein energy-momentum 
complex obtained the energy distribution of the Dymnikova space-time 
that is positive everywhere and be equal to zero at origin.  Later, 
Radinschi~\cite{11} calculate the energy distribution in the 
same one by using the Tolman energy-momentum complex and get the 
same result.  In this article, we would like to evaluate the energy 
distributions of the Dymnikova space-time by using the 
Weinberg~\cite{2}, Papapetrou~\cite{3} and M{\o}ller~\cite{4} 
energy-momentum complex.

In 1992, Dymnikova~\cite{12} obtained the static spherically 
symmetric nonsingular black hole solution, which behaves like the 
Schwarzschild solution but with its singularity replaced by the 
de-Sitter core.  The line element of solution is expressed as
\begin{equation}
ds^2 = \left( 1- \frac{R_g (r)}{r} \right) dt^2 - \left( 1- 
\frac{R_g (r)}{r} \right)^{-1} dr^2 - r^2 d\theta^2 
-r^2 \sin^2\theta d\phi^2 ,
\end{equation}
where 
\begin{equation}
  R_g (r) = r_g  \left( 1 - \exp (\frac{r^3}{r_*^3}) \right)  .
\end{equation}
We have also $r_*^3 = r_g r_0^2$, $r_g = 2M$ and $r_0^2 = \frac{3}{8 \pi \varepsilon_0}$. 
As $ r \rightarrow \infty$, the line element of solution becomes 
the Schwarzschild solution, and 
as $ r \rightarrow 0$, the line element of solution becomes
the de Sitter solution with energy-momentum density
$ T_{\mu\nu} = \varepsilon_0 g_{\mu\nu} $.

First, let us consider the Weinberg energy-momentum complex~\cite{2}
\begin{equation}
Q^{\rho \nu \lambda} = \frac{1}{2} \left\{ \frac{\partial h^{\mu}_{\mu}}{\partial x_{\nu}} \eta^{\rho \lambda}
- \frac{\partial h^{\mu}_{\mu}}{\partial x_{\rho}} \eta^{\nu \lambda} - \frac{\partial h^{\mu \nu}}
{\partial x_{\mu}} \eta^{\rho \lambda} + \frac{\partial h^{\mu \rho}}{\partial x_{\mu}} \eta^{\nu \lambda}
+ \frac{\partial h^{\nu \lambda}}{\partial x_{\rho}} - \frac{\partial h^{\rho \lambda}}{\partial x_{\nu}}
\right\}
\end{equation}
where the $\eta_{\mu \nu}$ is the Minkowski metric and $h_{\mu \nu} = g_{\mu \nu} - \eta_{\mu \nu}$.
The energy component of the Weinberg energy-momentum complex is most conveniently calculated in the 
quasi-Cartesian coordinate $(t,x,y,z)$, and using Gauss's theorem gives
\begin{eqnarray}
E_{\rm W}(r) & = & -\frac{1}{8\pi} \oint Q^{i00} n_i r^2 d\Omega  \\
             & = & -\frac{1}{16 \pi} \oint (h_{ij,j} - h_{jj,i} ) n_i r^2 d\Omega  ,       
\end{eqnarray}
the integral begin taken over a large sphere of radius $r$, with $\vec{n}$ the outward normal and $d\Omega$
the differential solid angle; that is 
$n_i = x_i/r$, $r^2 = x_i x_i$, and $d\Omega = \sin \theta d\theta d\phi$.
Herein the Latin index takes values from 1 to 3.  In these  
coordinates, the line element (1) reads
\begin{equation}
h_{\mu \nu} = \left[ \begin{array}{cccc}
-R_g (r) /r & 0 & 0 & 0 \\
0 & x^2 \Psi /r^2 & xy \Psi /r^2 & xz \Psi /r^2 \\
0 & xy \Psi /r^2 & y^2 \Psi /r^2 & yz \Psi /r^2 \\
0 & xz \Psi /r^2 & yz \Psi /r^2 & z^2 \Psi /r^2 \\
\end{array} \right] ,
\end{equation}
where $\Psi = (1-\frac{r}{r-R_g (r)})$ .
The required nonvanishing components $ Q^{i00} $ of the 
Weinberg energy complex are easily shown to be
\begin{eqnarray}
  Q^{100} = \frac{x}{r^2}(1- \frac{r}{r-R_g (r)}) , \\
  Q^{200} = \frac{y}{r^2}(1- \frac{r}{r-R_g (r)}) , \\
  Q^{300} = \frac{z}{r^2}(1- \frac{r}{r-R_g (r)}) .   
\end{eqnarray}
Hence, the Weinberg energy complex within radius $ r $ reads  
\begin{equation}
  E_{\rm W}(r) = \frac{R_g (r)}{2}(1- \frac{R_g (r)}{r})^{-1} . 
\end{equation}

Second, we calculate the energy component of the Papapetrou 
energy-momentum complex~\cite{3} in quasi-Cartesian coordinate 
\begin{equation}
E_{\rm P} (r) = \frac{1}{16\pi} \int \frac{\partial B^{00i}}
{\partial x^i} d^3 x 
\end{equation}
with
\begin{equation}
B^{00i} = \frac{\partial}{\partial x^j} [\sqrt{-g}
(g^{ij} + \eta^{ij} g^{00})]  .
\end{equation}
Herein the Latin index also takes values from 1 to 3.  The required 
nonvanishing components of $B^{00i}$ in the calculation of the 
Papapetrou energy complex are the following
\begin{eqnarray}
B^{001} = \frac{x}{r^2} (R_g^{\prime} (r) +\frac{R_g (r)}{r}) -
\frac{x}{r} \left[ \frac{rR_g^{\prime} (r)-R_g (r)}{(r-R_g (r))^2} 
\right] \\
B^{002} = \frac{y}{r^2} (R_g^{\prime} (r) +\frac{R_g (r)}{r}) -
\frac{y}{r} \left[ \frac{rR_g^{\prime} (r)-R_g (r)}{(r-R_g (r))^2} 
\right] \\
B^{003} = \frac{z}{r^2} (R_g^{\prime} (r) +\frac{R_g (r)}{r}) -
\frac{z}{r} \left[ \frac{rR_g^{\prime} (r)-R_g (r)}{(r-R_g (r))^2} 
\right] ,
\end{eqnarray}
where $R_g^{\prime} (r) =\frac{3r^2}{r_0^2} \exp(-\frac{r^3}
{r_*^3})$.  So, the energy within radius $r$ in the Papapetrou 
prescription is given as
\begin{equation}
E_{\rm P} (r) = \frac{1}{4} (R_g (r) +rR_g^{\prime} (r)) +\frac{1}{4}
\left[ \frac{R_g (r)-rR_g^{\prime} (r)}{(1-\frac{R_g (r)}{r})^2} 
\right] .
\end{equation}

Finally, we evaluate the energy distribution by the M{\o}ller 
energy-momentum complex~\cite{4} in spherical coordinates 
\begin{equation}
E_{\rm M} (r) = \frac{1}{8\pi} \int \frac{\partial \chi_0^{0i}}
{\partial x^i} d^3x 
\end{equation}
with 
\begin{equation}
\chi_0^{0i} = \sqrt{-g} g^{0\beta} g^{i\alpha}
\left( \frac{\partial g_{0\alpha}}{\partial x^{\beta}} - 
\frac{\partial g_{0\beta}}{\partial x^{\alpha}} \right) .
\end{equation}
In which the Latin index also takes values from 1 to 3, but Greek 
indices run from 0 to 3.  
In his recent article about to analyze the M{\o}ller energy-momentum
complex, Lessner~\cite{13} concludes that it is a powerful representation
of energy and momentum in general relativity.  Thus, Yang, Lin and 
Hsu~\cite{14} employing the M{\o}ller energy-momentum complex obtained the 
energy distribution of the charged dialton black hole recently. 
Later, Xulu~\cite{15} also computes the energy distribution in the 
Kerr-Newman space-time using the M{\o}ller energy-momentum complex.
Notice the only nonvanishing component 
of M{\o}ller energy-momentum complex is 
\begin{equation}
\chi_0^{01} = \sin \theta \left[ r_g - 
r_g \exp(-\frac{r^3}{r_*^3}) -\frac{3r^3}{r_0^2} 
\exp(-\frac{r^3}{r_*^3}) \right]  ,
\end{equation}
and the M{\o}ller energy-momentum complex within radius $r$ is
\begin{equation}
E_{\rm M} (r) = M - M \exp(-\frac{r^3}{r_*^3}) - 
\frac{3r^3}{2r_0^2} \exp(-\frac{r^3}{r_*^3}) .
\end{equation}

In summary, the energy component of those three energy-momentum 
complexes reduce $M$ as $r \rightarrow \infty$, corresponding to
\begin{equation}
E_{\rm W}(r)|_{r \rightarrow \infty} =  E_{\rm P}(r)|_{r \rightarrow 
\infty} = E_{\rm M}(r)|_{r \rightarrow \infty} = M  .
\end{equation}
Therefore, the total energy is given by parameter $M$ which is the 
same as the ADM mass for this metric, and is independent on 
definitions of energy-momentum complex.  According to the Cooperstock
hypothesis~\cite{16}, the energy and momentum are confined to the region of 
non-vanishing energy-momentum tensor of matter and all non-gravitation
fields.  The standard formula for the mass of de Sitter-Schwarzschild
solution obtained by Dymnikova 
\begin{equation}
m(r) = \int_0^r T_0^0 d^3 x =  \frac{R_g (r)}{2}
\end{equation}
is positive everywhere, unless at the origin, supports this hypothesis.
About a decade age, Bondi~\cite{17} argued that a nonlocalizable form of 
energy is not allowed in relativity and therefore its location can in 
principle be found.  Due to this reason, we try to comparision with the 
results of Yang~\cite{10} 
\begin{equation}
E_{\rm E} (r) = \frac{r_g}{2} \left( 1 - \exp \left( -\frac{r^3}
{r_*^3} \right) \right)
\end{equation}
and of Radinschi~\cite{11} 
\begin{equation}
E_{\rm T} (r) = \frac{r_g}{2} \left( 1 - \exp \left( -\frac{r^3}
{r_*^3} \right) \right) ,
\end{equation}
we could be a conclusion that 
\begin{equation}
E_{\rm E} (r) = E_{\rm T} (r) = g_{00} E_{\rm W} (r) = 
\frac{R_g (r)}{2} \equiv E_{\rm ETW} (r) 
\end{equation}
while Dymnikova space-time is expressed by Schwarzschild Cartesian 
coordinates, but $E_{\rm P} (r)$ and $E_{\rm M} (r)$ are both not 
equal to $E_{\rm ETW} (r)$. In other words, these results should be 
support to the idea that the Einstein, Tolman and Weinberg 
energy-momentum complexes can give the same result for a given 
space-time.  Also, this results sustain the idea~\cite{18} that the 
Papapetrou energy-momentum complex do not "coincide" with the
Einstein, Tolman and Weinberg energy-momentum complexes in 
the Schwarzschild Cartesian coordinate.

\begin{center}
{\bf Acknowledgements}
\end{center}
I.-C. Yang thanks the National Science Council of the 
Republic of China for financial support under the contract number 
NSC 89-2112-M-006-030.

\end{document}